\documentclass[ reprint,  amsmath,amssymb,  aps, aip,rsi,     ]{revtex4-1}%
\usepackage{graphicx}
\usepackage{dcolumn}
\usepackage{bm}
\usepackage{amsmath}
\usepackage{amsfonts}
\usepackage{amssymb}%
\usepackage{inputenc}

\providecommand{\U}[1]{\protect\rule{.1in}{.1in}}

\newcommand{\OO}[0]{O$_{2}$ }

\begin{document}
\title{{Perpendicular magnetic anisotropy of an ultrathin Fe layer grown on NiO(001)}}

\author{Soki Kobayashi} 
\author{Hiroki Koizumi} 
\altaffiliation[Present address:]{Research Center for Magnetic and Spintronic Materials, National Institute for Materials Science (NIMS), Tsukuba 305-0047, Japan}
\affiliation{Department of Applied Physics, University of Tsukuba, Tsukuba 305-8573, Japan}

\author{Hideto Yanagihara}\email{yanagihara.hideto.fm@u.tsukuba.ac.jp}
\affiliation{Department of Applied Physics, University of Tsukuba, Tsukuba 305-8573, Japan}
\affiliation{Tsukuba Research Center for Energy Materials Science (TREMS), University of Tsukuba, Tsukuba 305-8573, Japan}

\author{Jun Okabayashi} 
\affiliation{Research Center for Spectrochemistry, The University of Tokyo,  Tokyo 113-0033, Japan}

\author{Takahiro Kondo} 
\affiliation{Department of Materials Science, University of Tsukuba, Tsukuba 305-8573, Japan}
\affiliation{Tsukuba Research Center for Energy Materials Science (TREMS), University of Tsukuba, Tsukuba 305-8573, Japan}

\author{Takahide Kubota} 
\altaffiliation[Present address:]{Department of Applied Physics, Tohoku University, Sendai 980-0845, Japan}
\affiliation{Institute for Materials Research, Tohoku University, Sendai 980-8576, Japan}

\author{Koki Takanashi} 
\altaffiliation[Present address:]{Advanced Science Research Center, Japan Atomic Energy Agency, Tokai 319-1195, Japan.}
\affiliation{Institute for Materials Research, Tohoku University, Sendai 980-8576, Japan}

\author{Yoshiaki Sonobe} 
\affiliation{Research Organization for Nano \&  Life Innovation, Waseda University, Tokyo 162-0041, Japan}

\date{April. 29, 2023}

\begin{abstract}
The magnetic anisotropy and magnetic interactions at the interface between Fe and NiO(001) were investigated.
Depending on the growth conditions of the NiO(001) layers and the post-annealing temperature, the preferential magnetization direction of the ultrathin Fe layer grown on a NiO(001) layer changed from in-plane to a direction perpendicular to the film plane. 
The lattice constant of the NiO(001) layers parallel to the growth direction  increased with ${O_2}$ flow rate, while that parallel to the in-plane were locked onto the MgO(001) substrate regardless of the growth conditions of the NiO layers. 
Moreover, perpendicular magnetization was observed only when the NiO layer was grown with  ${O_2}$ flow rates higher than 2.0 sccm  corresponding to oxygen-rich NiO. 
X-ray magnetic circular dichroism  measurements revealed an enhancement in anisotropic orbital magnetic moments similar to the origin of perpendicular magnetic anisotropy at the Fe/MgO(001) interface. 
The interfacial magnetic anisotropy energies were 0.93 and 1.02 mJ/m$^2$ at room temperature and at 100 K, respectively, indicating less temperature dependence. 
In contrast, the coercivity $H_c$ exhibited a significant temperature dependence. Although no signature of exchange bias or unidirectional loop shift was observed, $H_c$ was strongly dependent on  the NiO layer thickness, indicating that the exchange interaction at the interface between the ferromagnetic and antiferromagnetic layers was not negligible, despite the NiO(001) being a spin-compensated surface. 
\end{abstract}

\maketitle
\section{Introduction}
Magnetic thin films with perpendicular magnetic anisotropy (PMA) are key components of spintronic devices to achieve high thermal stability and low switching current for magnetization reversal, which are crucial for realizing the high density and the low switching power magneto-resistive random-access memory\cite{{Bhatti2017SpintronicsReview}, {Sbiaa2011MaterialsMemory}}. 
Perpendicular magnetization films have been realized in various systems, such as magnetic compounds/alloys with relatively high uniaxial magneto-crystalline anisotropy\cite{Weller2000HighGbits/in2, Piramanayagam2007a,Shima2007HardFilms, Coey2011HardPerspective,  Mizukami2016Mn-basedSpintronics} in the form of thin films and magnetic multilayers\cite{Heinrich1993UltrathinInteractions,Johnson1996MagneticMultilayers, Miura2022UnderstandingMoments}.
The origin of PMA can be  divided into two mechanisms: a bulk effect, such as magneto-crystalline anisotropy\cite{Ravindran2001LargeCalculations, Sander2004TheFilms}, and an interfacial effect\cite{Dieny2017PerpendicularApplications}. Both magneto-crystalline anisotropy and interfacial magnetic anisotropy originate from spin-orbit interaction (SOI) associated with a crystal symmetry lowering. The interfacial magnetic anisotropy is particularly useful for controlling the preferential directions of the magnetic layers in spintronic devices owing to the stacking structures of various thin films.

Since the discovery of interfacial PMA in the bilayer system of CoFeB/MgO\cite{Yakata2009InfluenceJunctions}, followed by the excellent demonstration of magnetic tunneling junctions (MTSs)\cite{ Ikeda2010AJunction}, extensive research has been conducted on the enhancement of the PMA and voltage-controlled magnetic anisotropy (VCMA) in Fe/MgO\cite{Maruyama2009, Niranjan2010ElectricInterface}  and related materials\cite{Lambert2013QuantifyingInterface, Iida2018PerpendicularLayers, Xiang2018LargeHeterostructures, Nozaki2022EnhancingInterface}.
For a Fe/MgO system, {\it ab initio} calculations have shown that the interfacial PMA originates from the hybridization between Fe-$3d_{z^2}$ and O-$2p_z$ because of SOI \cite{Nakamura2010RoleInterface, Masuda2018PerpendicularFe/MgO, Miura2022UnderstandingMoments}.
Through X-ray magnetic circular dichroism (XMCD) measurements, Okabayashi \emph{et al.} showed that the origin of PMA at the interface can be attributed to the enhancement in anisotropic orbital angular momentum (OAM) of Fe induced by SOI\cite{Okabayashi2014PerpendicularDichroism, Okabayashi2019PerpendicularDichroisms}. 
Therefore, if a similar orbital hybridization is realized, other PMA systems may be observed at the interface between Fe and certain oxides, particularly isostructural oxides, such as MgO.

NiO is a typical antiferromagnetic material with a simple rock-salt structure and a N\'{e}el temperature of 523 K, which is sufficiently higher than room temperature. The lattice constant is close to those of nonmagnetic materials, such as Ag and MgO; therefore, NiO has been a typical antiferromagnetic compound for studying antiferromagnetic spintronics\cite{Moriyama2018SpinNiO, Moriyama2020EnhancedNiO/Pd, Chirac2020UltrafastTorques, Yang2011NegativeBarriers}.  Kozio\l-Rachwa\l \ \emph{et al.} recently found that the preferred magnetization direction of the Fe layer of  Fe/NiO/MgO(001) in-plane and that the N\'{e}el vector of NiO changes from  out-of-plane to in-plane if a Cr layer is inserted between the NiO layers and MgO(001) substrates owing to the change in the in-plane lattice constant\cite{Kozio-Rachwa2020ControlCoupling}.
Thus, Fe/NiO is a bilayer composed of the conventional ferromagnet and antiferromagnet and also comprises a fascinating interface in terms of the interplay and  cooperation between the two magnetic layers with different magnetisms.

In this study, we demonstrated that ultrathin Fe(001) becomes a perpendicular magnetization film owing to the interfacial PMA at the interface between Fe(001) and off-stoichiometric NiO(001). 
The angular dependence of XMCD measurement revealed that the origin of PMA is OAM of Fe, which is similar to those of previously reported Fe/MgO systems. 
 
 The growth of atomically flat Fe(001) layers with a few monolayers is believed to be  crucial for enhancing the interfacial PMA in a Fe/MgO(001) system. Because of the low wettability of the Fe film on MgO, most of the previously reported perpendicular magnetization films were realized at the interface between the top MgO(001) and bottom Fe(001) layers. Thus, it is challenging to achieve interfacial PMA with the reverse-stacking structure\cite{Nozaki2013GrowthLayerb}.

\section{Experiment}\label{sec2}
All the samples were grown on single crystal MgO(001) substrates using a  radio-frequency magnetron sputtering technique. 
The stacking structure reported in this study was Cr(001)/Fe(001) /NiO(001) /MgO(001)(substrate). Nickel oxide layers were grown at 500 $^{\circ}$C using a metal Ni target in a mixture of Ar and O$_2$. Successively, Fe and Cr layers were grown at room temperature using the same sputtering system. Following deposition, the samples were annealed for 1 h in vacuum as a post annealing process. The flow rate of Ar was fixed at 10 sccm.

\begin{figure}[ht]
\begin{center} 
\includegraphics[keepaspectratio , width=8cm]{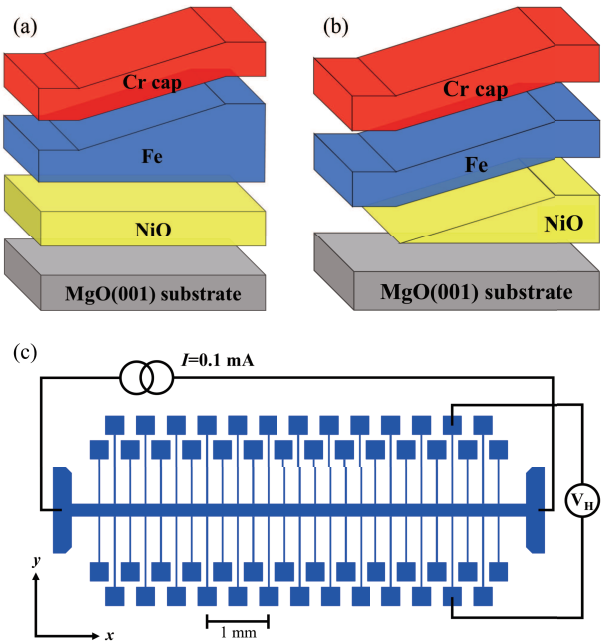}
\caption{(a) and (b) Stacking structures of Cr(2 nm)/Fe/NiO/MgO(001) substrates with wedge-shaped thickness gradients. (c) Hall pattern and wiring arrangement. The current path is parallel to the gradient direction of the wedge.}
\label{Hall1}
\end{center} 
\end{figure}

We prepared four types of samples for optimizing the PMA, quantitatively analyzing the PMA, and examining the magnetic interaction between ferromagnetic Fe and NiO(001) layers. 
The first type of sample was a series of multilayers of Cr(2 nm)/Fe(1 nm)/NiO(20 nm)/MgO(001), where the NiO(001) layers were grown under \OO flow rates in the range of 0.5-6.0 sccm. Hereinafter, the length in parentheses of the stacking structure expresses the thickness of each layer.  The samples were subsequently heated at 350$^{\circ}$C in vacuum for 1 h. The \OO flow rate was used as a growth parameter.
The second sample was a series of multilayers of Cr(2 nm)/Fe(1 nm)/NiO(20 nm) /MgO(001) subjected to different post-annealing temperatures to optimize the PMA. The \OO flow rate was fixed at 2.0 sccm. The third and final samples were  multilayers of Cr(2 nm)/Fe/NiO(20 nm)/MgO(001) with a wedge-shaped Fe layer (0.5-4.0 nm) (Fig. \ref{Hall1}(a)) and Cr(2 nm)/Fe (0.6 nm) /NiO/MgO(001) with a wedge-shaped NiO layer (0 - 30 nm)  (Fig. \ref{Hall1}(b)), respectively, for anomalous Hall effect (AHE) measurements. Both the samples were grown under \OO flow rate of 2.0 sccm and post-annealing temperature of 350$^{\circ}$C. Wedge-shaped layers were prepared using a linear moving mask. 
In addition, to evaluate the dependence of the NiO(001) film structures and valence states of Ni on the reactive sputtering process,  NiO/MgO(001) films were prepared at various \OO flow rates and process temperatures without Fe or Cr layers.

The samples were characterized using reflection high-energy electron diffraction (RHEED), X-ray reflectivity (XRR), X-ray diffraction (XRD), reciprocal space mapping (RSM), and X-ray photoelectron spectroscopy (XPS). XRR, XRD, and RSM experiments were performed using a Rigaku SmartLab with an X-ray source of Co-$K\alpha_{1}$.
The  magnetization of the multilayer films  was measured using a vibrating sample magnetometer (VSM) at room temperature. 
In addition, XMCD and X-ray absorption spectroscopy (XAS) measurements  were conducted at BL-7A, Photon Factory, high-energy accelerator organization KEK-PF. A magnetic field of $\mu_0 H = \pm1.2$ T was applied along the incident polarized beam by switching the magnetic field directions. The total electron yield mode was adopted. The geometry between the sample surface normal and incident beam directions was varied by changing the sample position from normal incidence (NI) to a grazing incidence of 60$^\circ$ (GI). All the XMCD measurements were performed at room temperature.
Subsequently, to evaluate the magnetization process dependence on both the Fe- and NiO-layer thicknesses, we performed AHE measurements on wedged-shaped films.  
The samples for AHE measurements were patterned into a Hall bar with many voltage probes on films of different thicknesses, as shown in Fig. \ref{Hall1} (c),  using photolithography and Ar ion milling. Cr (10 nm) and Au (100 nm) were sputtered on the electrical contact pads. The current path was parallel to the direction of the gradient of the film with a wedge-shaped thickness distribution, and the Hall voltages were measured at positions with different Fe or NiO thicknesses. The typical applied current was 0.1 mA.
All the measurements were performed at room temperature, unless otherwise stated.

\section{Results and Discussion}\label{sec3}

\subsection{Epitaxial growth of NiO(001) films}\label{subsecA}
\begin{figure}[ht]
\begin{center} 
\includegraphics[keepaspectratio , width=8cm]{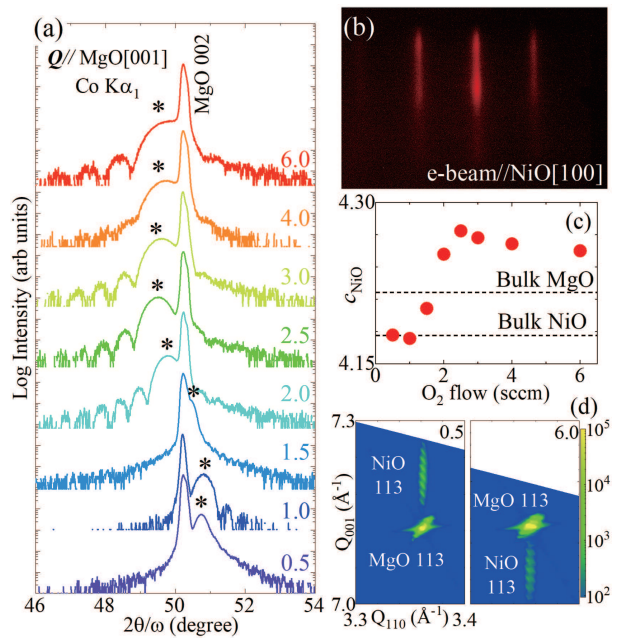}
\caption{(a) $2\theta/\omega$-XRD patterns of NiO(001) films grown under various different \OO flow rates. The asterisks indicate 002 peaks of NiO.  (b) Typical RHEED pattern of NiO(001). (c) \OO flow rate dependence of the lattice constants along the growth direction ($c_{\mathrm{NiO}}$) determined by $2\theta/\omega$-XRD measurements. (d) RSMs around the 113 diffraction of NiO/MgO(001) films grown at \OO flow rates of 0.5 sccm (left) and 6 sccm (right). The numbers shown on the right side in (a) and (d) indicate the oxygen flow rates during deposition.}
\label{C-O2}
\end{center} 
\end{figure}

First, we investigated the valence states and composition, as well as the lattice parameters of NiO, depending on the \OO flow rate during the growth processes.  As shown in Fig. \ref{C-O2}(a), clear Laue fringes are observed around the 002 diffraction of the XRD patterns, indicating that the NiO films are epitaxially grown with a (001) orientation and had a sufficiently smooth surface without incoherent lattice distortion at any \OO flow rate. In addition, the RHEED images of all  samples exhibited typical streak patterns (Fig. \ref{C-O2}(b)),  implying that the film surfaces are atomically flat and barely distorted, which is consistent with the observation of the Laue fringes.
Figure \ref{C-O2}(c) summarizes the lattice constants normal to the film plane of $c_{\mathrm{NiO}}$ as a function of the \OO flow rate, as determined by 002 reflection positions of the XRD patterns.
$c_{\mathrm{NiO}}$ is largely the same as that of the bulk value of NiO (4.176 \AA)\cite{Navrotsky1973ThermodynamicCoOMgOGeO2} for \OO flow rates $\leq$ 1.0 sccm, and $c_{\mathrm{NiO}}$ becomes greater than that of the bulk value of MgO (4.216 \AA) for \OO flow $\geq$ 2.0 sccm, as shown in Fig. \ref{C-O2}(c). 

As shown in Fig. \ref{C-O2}(d), the RSM measurements indicate that the lattice constants along the in-plane direction of NiO(001) mostly matched that of MgO of the substrate at any \OO flow rate, implying  $a_{\mathrm{NiO}} \approx $4.22 \AA. 
Moreover, clear fringe patterns were observed around NiO 113 in both the films, suggesting that all the NiO films were coherently distorted owing to epitaxial stress. Moreover, the critical thickness of the misfit relaxation was greater than 20 nm for all the films.
Because the in-plane lattice constants are locked to the MgO(001) substrate, the volume of the unit cell varies with the \OO flow. 

Similar results have been reported in previous studies \cite{Chen2010MicrostructuresSputtering, Wang2017TheFilms, Ferreira1996ElectrochromicConditions}.   
With increasing oxygen in the film growth process,  Ni$^{2+}$ ions are replaced by Ni$^{3+}$ ions and vacancies at the Ni-sites, resulting in the growth of non-stoichiometric NiO(001) films.
We also performed XPS measurements of the NiO(20 nm)/MgO(001) films without Fe or Cr layers grown at various \OO flow rates. The relative composition of oxygen and a trace of Ni$^{3+}$ in NiO increased with the increase in the \OO flow rate,  consistent with  previous XPS measurements\cite{Oswald2004XPSFilms} and the  structural analysis mentioned.  

\subsection{Interface magnetic anisotropy of Fe/NiO(001)}\label{subsecB}
\begin{figure}
\begin{center} 
\includegraphics[width=8cm]{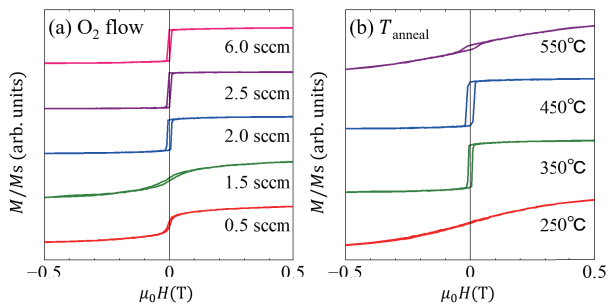}
\caption{Out-of-plane $MH$ loops of Cr/Fe/NiO(001) multilayers at room temperature; (a) \OO flow dependence range of 0.5 to 6.0 sccm with $T_{\rm{anneal}}=$500$^{\circ}$C  and  (b)$T_{\rm{anneal}}$ dependence range of 250 to 550$^{\circ}$C with \OO flow rate of 2.0 sccm.}
\end{center} 
\end{figure}
To study the magnetic anisotropy at the Fe/NiO(001) interface, we fabricated Fe/NiO multilayers on MgO(001) substrates with different \OO flow rates for NiO(001) layer growth. 
Figure \ref{VSM1}(a) shows the out-of-plane magnetization processes of 0.63 nm-thick Fe thin films grown on 20 nm-thick NiO(001) layers. 
Clearly, the magnetization processes are sensitive to the growth conditions of NiO(001), and the Fe layer becomes perpendicular magnetization. Therefore, PMA was dominant when the \OO flow rate $\geq$ 2.0 sccm. 
The in-plane lattice constant of NiO(001) was largely the same as that of the MgO(001) substrate, irrespective of the growth condition of the NiO layer, as mentioned in  Sec. \ref{subsecA}. The observed PMA dependence on the growth condition of the NiO(001) layer originates from an interfacial effect rather than an epitaxial strain.
Next, we fabricated a Cr/Fe/NiO (\OO flow rate = 2.0 sccm) structure employing a post-annealing process at different temperatures ($T_{\rm{anneal}}$) following the growth of the capping layer of Cr. 
Figure \ref{VSM1}(b) shows the out-of-plane magnetization processes of the films annealed at different $T_{\rm{anneal}}$. 
The samples were perpendicular magnetization films at  $T_{\rm{anneal}}$ = 350 and 450$^{\circ}$C . 
In the case of $T_{\rm{anneal}}$ = 550$^{\circ}$C, a clear peak shift of NiO 002 was observed in XRD patterns (not shown), suggesting considerable atomic mixing at the interface.

We performed AHE measurements on  multilayer Cr-cap (2 nm)/Fe (0.5-4.0 nm)/NiO (20 nm)/MgO(001)(substrate) to quantitatively separate the observed magnetic anisotropy into the interfacial magnetic anisotropy and volume contribution. 
The sample with the Fe layer thickness of $t_{\rm{Fe}}$  = 0.5 nm  is perpendicular magnetization.
In contrast, the saturation field increased with increase in $t_{Fe}$, indicating that the shape anisotropy became dominant, and that the preferential direction of the magnetization  changed from normal to the film plane and finally to the in-plane direction. 
These results indicate that the origin of PMA is the interfacial effect rather than the bulk effect.

\begin{figure}
\begin{center} 
\includegraphics[width=8cm]{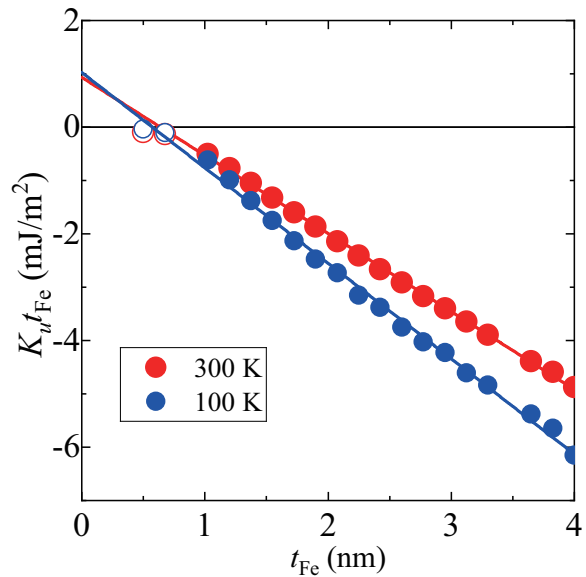}
\caption{
$K_u t_{\rm{Fe}}$-$ t_{\rm{Fe}}$ plot at 100 K (blue circles) and room temperature (red circles). $K_v$ and $K_i$ are determined from the slope and intercept, respectively. The thinnest two data points expressed as open circles are excluded from the fits due to the perpendicular magnetization samples.}
\label{Kut2}
\end{center} 
\end{figure}

The areal magnetic anisotropy ($K_u t_{\rm{Fe}}$) as a function of $t_{\rm{Fe}}$, $K_u t_{\rm{Fe}} = K_v t_{\rm{Fe}} + K_i$ \cite{denBroeder1991MagneticMultilayers}, is plotted in Fig. \ref{Kut2} . Here, $K_v$ is the volume contribution to magnetic anisotropy, for example, magneto-crystalline, strain induced, and shape anisotropies, while $K_i$ is the interface contribution. 
$K_i$ at room temperature was determined as $0.93\pm 0.03$ mJ/m$^2$. This is of the same order as the previously reported $K_i$ for the Fe/MgO interface \cite{Koo2013LargeInterface,Okabayashi2014PerpendicularDichroism}.
In contrast, $K_v$ was estimated to be $-1.46 \pm0.01$ MJ/m$^3$, which is reasonably close to the demagnetization energy or shape anisotropy energy for the film form a sample with $-\frac1 2 \mu_0M_s^2 = -1.23$ MJ/m$^3$. Here, $M_s$ denotes the saturation magnetization of Fe, which are 1400 and 1700 kA/m at room temperature and 100 K, respectively as determined by VSM.  We also performed the same analysis at 100 K, and  obtained $K_i$ and $K_v$ are $+1.02\pm 0.07$ mJ/m$^2$ and $-1.79\pm 0.03$ MJ/m$^3$, respectively. 

We emphasize that the temperature dependence of the interfacial PMA of Fe/NiO appears to be weaker than that of CoFeB/MgO\cite{Sato2018Temperature-dependentSimulations}. As the  PMA is dominated by magnetization at the interface of Fe layer\cite{Ibrahim2020UnveilingInterfaces},  the observed weaker temperature dependence of the PMA in Fe/NiO compared to CoFeB/MgO indicates that the magnetic exchange coupling between Fe and NiO at the interface may suppress thermal fluctuations of the Fe spins. Therefore, this system could be useful in applications owing to the greater robustness against thermal fluctuations.

\subsection{XAS and XMCD analyses}\label{subsecC}
\begin{figure} 
\begin{center} 
\includegraphics[width=9cm]{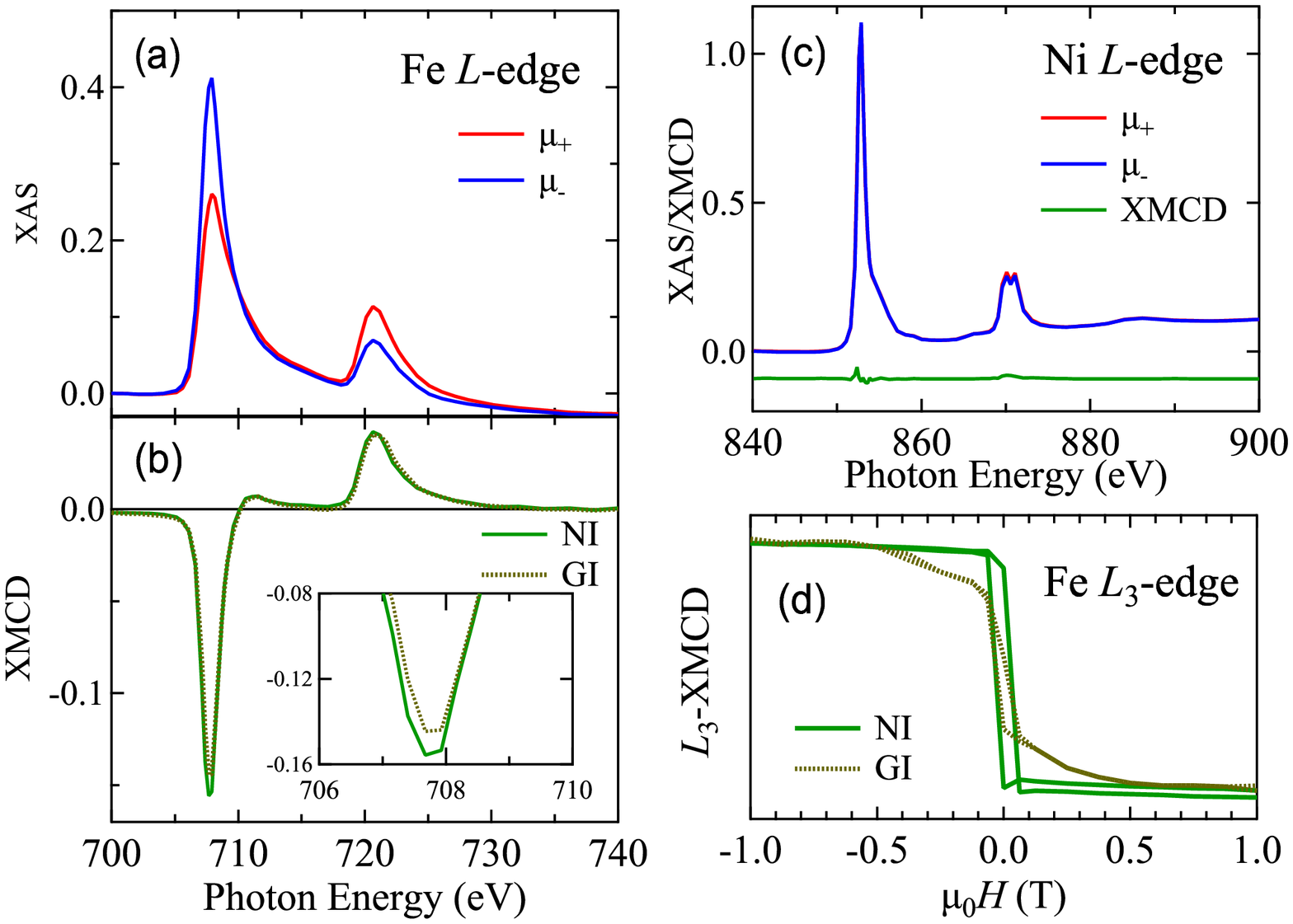}
\caption{XAS and XMCD of the perpendicular magnetization film of Fe/NiO(001). (a) XAS in Fe $L$-edge along the different magnetic field directions with $\mu_+$ and $\mu_-$, in the NI setup.  (b) XMCD ($\mu_+-\mu_-$) for the NI and GI geometries. The inset shows an expanded view around the $L_3$-edge.  (c) XAS and XMCD spectra at the Ni $L$-edge. (d) XMCD hysteresis curves in NI and GI geometries. 
}
\label{xmcdPMA}
\end{center} 
\end{figure}
\begin{figure}
\begin{center} 
\includegraphics[
width=9cm]{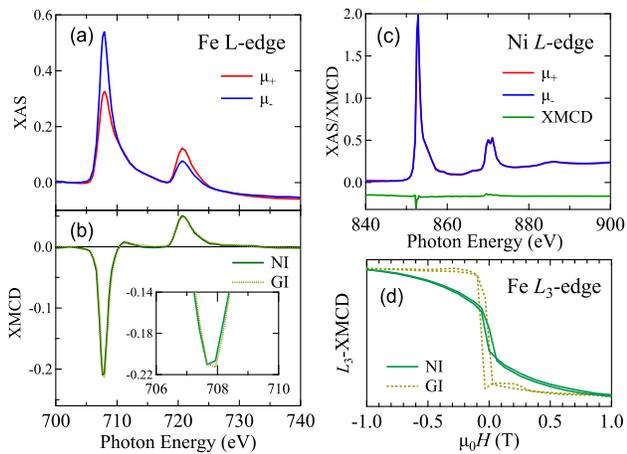}
\caption{XAS and XMCD of the in-plane magnetization film of Fe/NiO(001). (a) XAS in Fe $L$-edge along the different magnetic field directions with $\mu_+$ and $\mu_-$, in the NI setup. (b) XMCD ($\mu_+-\mu_-$) for the NI and GI geometries. The inset shows an expanded view around the $L_3$-edge.  (c) XAS and XMCD spectra at the Ni $L$-edge. (d) XMCD hysteresis curves in NI and GI geometries. 
}
\label{xmcdIMA}
\end{center} 
\end{figure}

Figures \ref{xmcdPMA} and \ref{xmcdIMA} show the Fe and Ni $L$-edge X-ray absorption spectra as well as the XMCD with angular dependence of the normal incidence (NI) and grazing incidence (60$^\circ$) (GI) cases of Cr(2 nm)/Fe(0.6 nm)/NiO(20 nm)/MgO(001), respectively. The NiO(001) layers were grown at \OO flow rates of 2.5 and 1.5 sccm for the perpendicular magnetization  (Fig. \ref{xmcdPMA}) and in-plane magnetization (Fig. \ref{xmcdIMA}) samples, respectively. For both the samples, the clearly observed metallic line shapes of Fe suggest a lack of interfacial oxidation. 
In the case of the perpendicular magnetization sample, the XMCD intensity at the $L_3$-edge varied with the measurement geometry. 
The enhanced $L_3$-peak in the NI geometry corresponded to orbital magnetic moments ($m_{\rm orb}$). By applying the sum rule analysis, we obtained the spin magnetic moments and $m_{\rm orb}$ to be 1.32 and 0.09 $\mu_B$, respectively, for the NI case. Here, the 3d hole number of 3.39 was assumed, which is the same as the case for Fe/MgO \cite{Okabayashi2019PerpendicularDichroisms}. If we express the $m_{\rm orb}$ components parallel and normal to the film plane as $m_{\rm orb}^{\perp}$ and $m_{\rm orb}^{\parallel}$, respectively, the difference of $m_{\rm orb}$, defined as $\Delta m_{\rm orb} = m_{\rm orb}^\perp - m_{\rm orb}^{\parallel}\simeq 0.05 \mu_B$, provides the reasonable origin of the observed PMA\cite{Miura2022UnderstandingMoments}.Here, we assumed that  the orbital moment anisotropy originated from the second order perturbation of the spin-orbit interaction\cite{Miura2022UnderstandingMoments}. A magnetic dipole moment of less than 0.01  $\mu_B$ was also estimated, which is sufficiently smaller than $m_{\rm orb}$. This suggests that the orbital magnetic moment anisotropy is dominant at the Fe/NiO interface.  In contrast, in the case of the in-plane magnetization sample, $L_3$-edge intensity remained unchanged for NI and GI set up, which suggests the isotropic $m_{\rm orb}$ of 0.05 $\mu_B$ because the shape anisotropy becomes dominant.  Hysteresis loops at Fe $L_3$-edge photon energy shown in Figs.\ref{xmcdPMA}(d) and \ref{xmcdIMA}(d) roughly reproduce the characteristic magnetization processes for the perpendicular  and in-plane magnetization films, respectively. The asymmetry in XMCD spectrum reveals that the large orbital moments are induced in the perpendicular components. These features are quite similar to the case of Fe/MgO\cite{Okabayashi2014PerpendicularDichroism}. However, the effect of Fe-O bonding to anisotropic charge distribution does not work in the in-plane magnetization sample even with 0.6 nm thickness.
As shown in Figs.\ref{xmcdPMA} and \ref{xmcdIMA}, there are no XMCD signals at the Ni $L$-edge for both the NI and GI cases because antiparallel spins at the Ni sites are compensated completely, although  small differential signals appeared, originating from application of the magnetic field along the perpendicular direction during the measurements.  

Interestingly, there were no signs of Ni$^{3+}$ in XAS, as shown in Figs. \ref{xmcdPMA}  and  \ref{xmcdIMA} (c). This result appears to be inconsistent with the XPS and XRD results, which indicate that the oxygen-rich NiO contains Ni$^{3+}$ and Ni-site vacancies in addition to Ni$^{2+}$ (Sec. \ref{subsecA}). The fact that no traces of iron oxide were detected (Fig. \ref{xmcdPMA} (a)) implies that no significant oxidation occurred between the excess oxygen in NiO and the metallic Fe layer. 
Koo \emph{et al.} pointed out that a variation in the interface composition due to oxygen atoms floating up from the Cr buffer layer and reaching the MgO interface is  key to enhancing the PMA in MgO/Fe/Cr/MgO(001)\cite{Koo2013LargeInterface}. Similarly, the variation or diffusion of excess oxygen in the NiO layer by annealing may play an important role in enhancing the PMA in the Fe/NiO system. Based on this, the XAS results of no signs of Ni$^{3+}$  can be consistently explained by the migration of excess oxygen in the NiO layer to the Cr layer which works as an oxygen absorber, and the interface structure of Fe/NiO becomes optimized to exhibit PMA.   

\subsection{Effects of coexistence of PMA and magnetic exchange at the interface of Fe and NiO}\label{subsecD}
\begin{figure}
\begin{center} 
\includegraphics[
width=8cm]{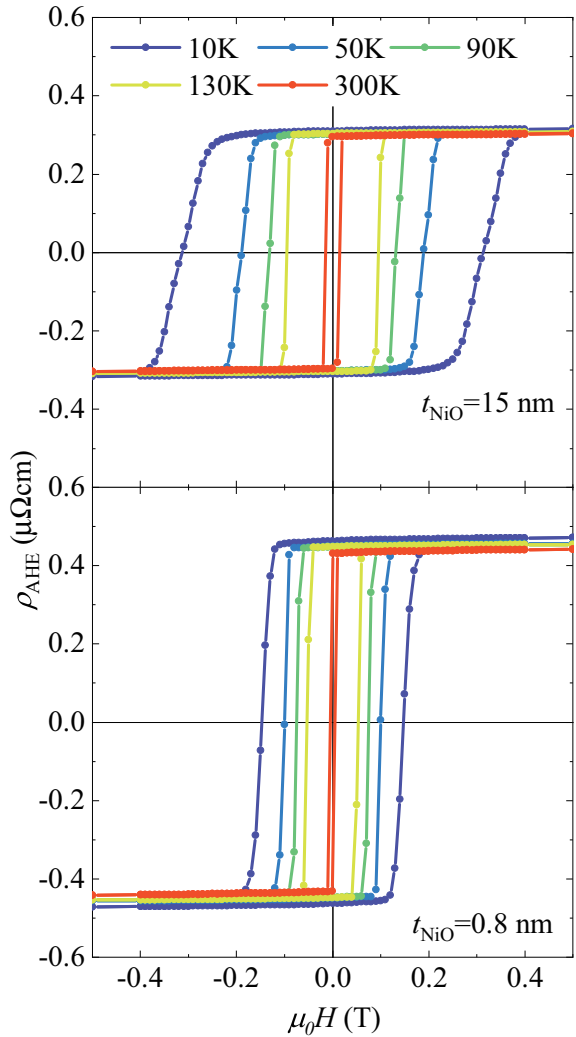}
\caption{$\rho _{\mathrm{AHE}}$ of Cr/Fe(0.6 nm)/NiO($t_{\mathrm {NiO}}$)/MgO(001) with $t_{\mathrm{NiO}}= 15$ nm (top) and $0.8$ nm (bottom). AHE measurements were performed at 10, 50, 90, 130, and 300 K. The thick-NiO sample of $t_{\mathrm{NiO}}=15$ nm exhibits larger $H_c$ at each temperature than the thin-NiO sample of $t_{\mathrm{NiO}}=0.8$ nm. }
\label{Hc_T}
\end{center} 
\end{figure}

To investigate the dependence of the NiO layer thickness  on the magnetization process, we conducted AHE measurements on the stacking of Cr/Fe(0.6 nm)/NiO($t_{\mathrm{NiO}}$)/MgO(001) with a wedge-shaped NiO layer ($t_{\mathrm{NiO}}=$ 0 - 30 nm)  (Fig. \ref{Hall1}(b)). 
As shown in Fig. \ref{Hc_T}, the films with thicker NiO layers have a higher coercive force $H_c$ at all temperatures and a greater temperature dependence of $H_c$. 
Generally, magnetic anisotropy is among the dominant parameters determining $H_c$; However, as mentioned in Sec. \ref{subsecB}, the PMA exhibited little temperature change. The fact that the temperature dependence of $H_c$ is more significant than that of PMA means that $H_c$ is dominated  by the bulky features of the NiO layer rather than the interfacial effect. The difference between the two samples with different NiO thicknesses could be explained by considering the blocking temperature ($T_B$) of antiferromagnets\cite{Meiklejohn1957,Ambrose1998DependenceBilayers,Devasahayam1999TheConditions}.  
According to the previous study of $T_B $ in antiferromagnets by Devasahayam \emph{et al.}\cite{Devasahayam1999TheConditions}, the thickness dependence of  $T_B$  can be expressed by the power law owing to the finite-size effect. The estimated correlation length was $\approx 2$ nm for NiO films and therefore, $T_B$ approaches or becomes less than the room temperature if $t_{\mathrm{NiO}} \lesssim 2$ nm. Thus, our sample with $t_{\mathrm{NiO}} = 15 $ nm has $T_B$  above 300 K,  while in the case of the sample with $t_{\mathrm{NiO}} = 0.8 $ nm, the $T_B$ is well below room temperature. As $t_{\mathrm{NiO}}$ of the latter sample is sufficiently thinner than the correlation length of NiO, the role of the exchange coupling between NiO and Fe at the interface or magnetic effect of antiferromagnetism of the NiO layer must be negligibly small. In contrast, the magnetization process of Fe layer in the the sample with $t_{\mathrm{NiO}} = 15 $ nm is strongly affected by the antiferromagnetic nature of NiO through the interfacial exchange coupling.

We also examined whether exchange bias emerges at the interface between the ferromagnetic Fe(001) layer and the antiferromagnetic NiO(001) layer for samples with an additional field annealing process. 
The details of the annealing process are as follows. Three films of Cr/Fe(0.6 nm)/NiO(20 nm)  were simultaneously grown on MgO(001) substrates with dimensions of 10 mm $\times$ 10 mm $\times$ 0.5 mm. Two of the three films were annealed at 250$^{\circ}$C in vacuum with a magnetic field of $\mu_0 H =$1 T for 1 h. They were placed parallel and perpendicular to the magnetic field, respectively. Subsequently, the samples were spontaneously cooled to room temperature under the magnetic field. The third sample was annealed in zero magnetic field as a control sample.

Hysteresis loop shifts, which are characteristic of the exchange bias effect, have often been reported in Fe/NiO(001), which exhibits in-plane preferential magnetization\cite{Luches2010MagneticLayers, Luches2012Depth-dependentBilayers,Mynczak2013NiO/Fe001:Structure,Li2019ChiralityAnisotropy,Kozio-Rachwa2020ControlCoupling} rather than perpendicular magnetization. Even at an interface between a ferromagnet and a compensated antiferromagnet layers, a finite exchange bias could be realized if the spins of both the ferromagnet and antiferromagnet lie in-plane because of the frustration of the first antiferromagetic layer\cite{Kiwi1999Exchange-biasInterfaces}.
In our case, no clear shift in the magnetization curve or a clear change in $H_c$ was observed in all three samples, regardless of the presence of the applied field during the annealing process or the direction of the applied field. This means that the antiferromagnetic spin configuration near the interface cannot be controlled by an external field, although the origin of $H_c$ could be dominated by interfacial spin frustration\cite{REF:Nogues1999, Leighton2000CoercivityFrustration}.
We would like to emphasize that the difference between whether Fe is a perpendicularly magnetized film or an in-plane magnetized film appears to govern the absence or presence of an exchange bias in the Fe/NiO(001) system.
Moreover, the relationship between the N\'eel vector of NiO(001) and the preferential magnetization direction of the ferromagnetic layer is  important for understanding the exchange bias mechanism in such bilayer systems.

\section{Conclusions}
In summary, we investigated the interfacial magnetic anisotropy at the interface between Fe and Ni oxide. 
We found that PMA emerged at the Fe/NiO interface at high \OO flow rates for NiO layer growth.
By measuring the Fe-thickness dependence of $K_u$, we obtained  $K_i$  of 0.93 and 1.02 mJ/m$^2$ at room temperature and 100 K, respectively. Further, XMCD measurements revealed an enhancement in the anisotropic magnetic moment of Fe, which was also observed at the  Fe/MgO interface.  
In contranst ot the lower temperature dependence of $K_i$, $H_c$ was strongly enhanced with decreasing temperature. However, no hysteresis loop-shift, which is characteristic of the exchange bias effect, was observed. The VCMA effect of Fe/NiO(001) should be examined in the near future. 
In addition to the well-known bilayer Fe/MgO(001) system, which shows PMA, we found that Fe/NiO(001) also exhibits relatively strong PMA. Because a rock-salt structure is a common crystal structure for divalent oxides, the combination of Fe and rock-salt type oxides is a promising system to exhibit greater PMA. 

\begin{acknowledgments}
This work was partly supported by JSPS KAKENHI(19KK0104, 21H01750, and 22H04966), TIA-KAKEHASHI (TK22-023), and Cooperative Research Project Program of the Research Institute of Electrical Communication (RIEC), Tohoku University.
The XMCD experiments were performed under the approval of the "Photon Factory Program Advisory Committee" (proposal No. 2021G069). 
Part of this work was supported by the Advanced Research Infrastructure for Materials and Nanotechnology in Japan (ARIM). 
We thank Seiji Mitani, Hiroaki Sukegawa, and Yoshio Miura for providing us with useful suggestions. 
\end{acknowledgments}

\end{document}